\documentclass[reprint,amsmath,amssymb, aps,showpacs]{revtex4-1}
\usepackage[dvipdfmx]{graphicx}

\usepackage{dcolumn}% Align table columns on decimal point
\usepackage{bm}% bold math
\usepackage{color}

\usepackage{here}
\usepackage{subfigure}

\begin{document}

\preprint{\bf PREPRINT}

\title{Diffusion of interface and heat conduction in the three-dimensional Ising model}

\author{Yusuke Masumoto}
 \email{masumoto@scphys.kyoto-u.ac.jp}
\author{Shinji Takesue}
 \email{takesue@scphys.kyoto-u.ac.jp}
\affiliation{
Department of Physics, Kyoto University, Kyoto 6068502, Japan\\}
\date{\today}

\begin{abstract}
We investigate the relationship between a diffusive motion of an interface, 
heat conduction, and the roughening transition in the three-dimensional Ising model. 
We numerically compute the thermal conductivity and the diffusion constant and
find that the diffusion constant shows a crossover in its temperature dependence.
The crossover temperature is equal to the roughening transition temperature in equilibrium
and deviates from it when heat flows in the system. 
From these results, we discuss the possibility that 
heat conduction causes a shift of the roughening transition temperature. 
\end{abstract}

\maketitle

%\tableofcontents
\section{Introduction}

Dynamical properties of an interface are being 
studied in relation to various physical phenomena such as crystal growth\cite{1}, 
grain-boundary\cite{2} and magnetic domain wall. 
Recently, inspired by the discovery of the spin Seebeck effect, 
the interface motion induced by a temperature gradient was examined 
theoretically\cite{4} and experimentally\cite{3}. In both cases, an interface 
moves to the hotter part of the system. According to the theoretical studies using 
the stochastic Landau-Lifshitz-Gilbert and Landau-Lifshitz-Bloch equations\cite{4},
this is due to a magnonic spin current and the conservation of angular momentum. 
Assuming local equilibrium, however, we can explain the result using a 
thermodynamic argument\cite{4,20}. 
Let us define the free energy of an interface as the difference between 
the free energy of a system with an interface and that of the same system 
without an interface. It becomes a monotonically decreasing function of 
temperature and vanishes at $T_c$. Then the interface moves towards 
the hotter region to minimize the free energy. 
Since this explanation is quite general, it should be applicable to a wide range of 
systems with an interface.

In the previous paper\cite{6}, we also found that the interface motion 
in the two-dimensional Ising model is a diffusion process with a drift force 
towards the high-temperature side, when no magnetic field is applied to the bulk 
and heat flows in the system. 
The strength of the drift force is proportional to the difference of temperature 
values at the two ends. 
Under an appropriate boundary condition, we prepared the system with an interface
and calculated the power spectrum of the temporal sequence of column-averaged
magnetizations.
It is known that when the step of the step function executes a random walk, 
the power spectrum of function values at a fixed position shows 
characteristic power-law behavior with exponent $-3/2$\cite{16}. 
In equilibrium states of the Ising model with an interface, the column-averaged 
magnetization shows such a power spectrum with some modification due to the  
finite width of the interface. To simulate heat conduction in the Ising model, 
we equipped the model with a cellular-automaton type energy-conserving dynamics.
We analytically calculated the power spectra in the case where an interface of a width
carries out diffusion with a drift.  
The obtained spectrum showed an excellent agreement with numerical results for
the heat conduction states. The thermodynamic explanation can be applied to our case, though not stated 
in the paper\cite{6}. We calculate the interface energy $\Delta E$ as the difference 
between the system energy under the antiparallel boundary condition minus that 
under the parallel one. Then the drag force estimated from the interface 
free energy $\Delta F(\beta)=\beta^{-1}\int_{0}^{\beta}\Delta E d\beta$ 
agrees with the numerical results.

To extend our research to three dimensions, we must consider possible influences 
from the roughening transition. The roughening transition is a phenomenon 
that a smooth surface turns into a rough one above a certain temperature 
called the roughening transition temperature. 
In the Ising model on the simple cubic lattice with isotropic couplings, the roughening
transition temperature is $T_{\mathrm{R}}^{\mathrm{eq}}=0.542 T_c$ 
\cite{8}, where $T_c$ is the critical temperature.
At a temperature higher than $T_{\mathrm{R}}^{\mathrm{eq}}$, 
the interface width is proportional to $\log L$,
where $L$ denotes the system size\cite{8}. It diverges in the thermodynamic limit.
In contrast, the interface width is constant for $L\gg 1$ at a temperature 
lower than $T_{\mathrm{R}}^{\mathrm{eq}}$.
The roughness of an interface can affect its motion. 
Some experiments show that the speed of crystal growth remarkably 
decreases below the roughening transition temperature\cite{5,15}. 
Thus, it is probable that the diffusion of an interface in the three-dimensional 
Ising system also shows some changes at the temperature.

In this paper, we focus on how the thermal conductivity and 
the diffusion coefficient vary near $T_{\mathrm{R}}^{\mathrm{eq}}$ and 
how heat conduction affects their behavior. 
In equilibrium, the diffusion constant shows different temperature dependence 
above and below $T_{\mathrm{R}}^{\mathrm{eq}}$. 
It decreases more rapidly below $T_{\mathrm{R}}^{\mathrm{eq}}$.
For thermal conductivity, the results depend on the time evolution rules for 
simulations. 
Thus, we employ two kinds of dynamics and compare those results. 
Moreover, we examine two kinds of arrangements of an interface. 
One is an interface perpendicular to heat flow and 
the other is an interface parallel to heat flow. 

Most interesting in our results is that when heat flows in the system, 
the crossover temperature at which the diffusion constant changes temperature 
dependence deviates from $T_{\mathrm{R}}^{\mathrm{eq}}$. 
It may indicate that the roughening transition temperature shifts 
in nonequilibrium situations. 

The paper is organized as follows. In Sec.~II,  we define the model and dynamics 
employed for simulations.  In Sec.~III we show simulation results on 
thermal conductivity.
In Sec.~IV, simulation results for diffusion coefficients are exhibited. 
Section V is devoted to summary and discussion.

\section{Setup of the system} 
In the literature, various kinds of spin dynamics have been proposed for the simulation 
of the Ising model. The most famous one is Glauber dynamics\cite{12}, 
where spins are stochastically updated according to some temperature-dependent 
transition rates. It is useful for investigating equilibrium properties of the Ising model 
because the detailed balance condition for the transition rates ensures 
that the system reaches an equilibrium state at the given temperature. 
However, it is not appropriate for the simulation of heat conduction, 
where local temperature values should be determined as a result of heat conduction. 

Creutz\cite{9} invented an alternative dynamics that conserves the following 
Hamiltonian.
\begin{equation}
H=-\sum_{<i,j>}\sigma_{i}\sigma_{j}+\sum_{i}4\tilde{\sigma}_{i}, 
\label{eq2}
\end{equation}
where $\sigma_{i}\in\{-1,+1\}$ denotes the Ising spin on site $i$ 
and $\tilde{\sigma}_{i}\in\{0,1,2,3\}$ is an auxiliary variable called ``momentum''.  
The first term means the usual ferromagnetic interaction and the second term 
is a kind of ``kinetic energy''. 
In each step, spin $\sigma_{i}$ is flipped if and only if the change in 
the interaction energy can be compensated by corresponding change of momentum 
variable $\tilde{\sigma}_{i}$. The condition is written as
\begin{equation}
0\leq\tilde{\sigma}_{i}-\frac{1}{2}\sigma_{i}(\sum_{nn}\sigma_{nn})\leq3,  \label{eq3}
\end{equation}
where $nn$ denotes the nearest-neighbor sites of $i$. 
If the above inequality is satisfied, spin $\sigma_{i}$ is flipped to $-\sigma_{i}$.
Because each momentum obeys the canonical distribution independently of each other,
local temperature values can be measured from the distributions or expectation
values of momentums. Creutz dynamics was successfully used in the study of 
heat conduction in the Ising model\cite{10}. 

A simplified variant of Creutz dynamics is Q2R\cite{17}, 
where the ``kinetic energy'' term is absent and spins are flipped only if the sum 
of the nearest-neighbor spins is zero.

In \cite{6}, we have found that Creutz dynamics has a serious problem at low temperature.
The interface motion becomes extremely slow and sometimes freezes.
Moreover, if the system is attached to a heat reservoir, 
it does not relax to the uniform equilibrium state at the reservoir temperature 
within simulation time.
Such problems arise from the following reasons.
Because most spins are in the same direction below $T_c$, 
a spin flip brings a large increase in the interaction energy. 
Although a large momentum is necessary to compensate it, 
it is rare at a low temperature. Thus, the dynamics becomes slow. 
By the same reason, the thermal conductivity shows a sudden drop 
around $T_c$\cite{10}.

The same problem is noticed in the Q2R and a solution to the problem
was brought by Casartelli et al \cite{14}. They combined a new dynamics called 
Kadanoff-Swift dynamics with the Q2R and called the resultant the KQ dynamics. 
In the KS dynamics,  a pair of 
next-nearest-neighbor spins exchanges the values 
if energy is unchanged by the exchange. 
It should be noted that the Hamiltonian does not have next-nearest-neighbor 
coupling terms. Such spins are only dynamically coupled. 
By employing the modified dynamics, relaxation to equilibrium was 
realized in simulation time.

Because we need to measure local temperature values in heat conduction, 
we modified Creutz dynamics in the similar manner by adding KS dynamics and
called the new dynamics Kadanoff-Swift-Creutz (KSC) dynamics in the study of
the two-dimensional Ising model\cite{6}. 
The KSC dynamics also realizes relaxation to equilibrium and 
the interface motion does not freeze in the two-dimensional systems.

Microcanonical (MC) dynamics is another spin dynamics that can be used 
at a low temperature\cite{11}. 
In the MC dynamics, the ``kinetic energy'' is defined not on each site but at each bond.
At each step, an update of randomly chosen two nearest-neighbor spins is considered.
We choose a candidate of new configurations for the two spins and calculate the
interaction energy variation. If it is compensated by the change of bond momentum, 
we accept the move. It was originally introduced to simulate a disordered system 
because the dynamics do not assume a regular lattice structure. 
It also shows the advantage of high mobility of energy even at a low temperature.
In the numerical study in this paper, we employ the KSC dynamics and the MC dynamics
and compare the results from the two dynamics.

To simulate heat conduction, the boundary spins in contact with heat reservoirs 
are evolved by Glauber dynamics. 
The temperature of left heat reservoir is denoted by $T_1$ and 
that of right heat reservoir is by $T_2(\leq T_1<T_c)$.
We also use average temperature $T=(T_1+T_2)/2$ and the temperature difference
$\Delta T=T_1-T_2$. 
Note that we employ energy unit where Boltzmann constant is unity and 
the critical temperature of the three-dimensional Ising model 
is $1/T_c=0.221654626(5)$ \cite{13}.
We can simulate heat conduction using deterministic energy-conserving dynamics 
such as the KSC dynamics or the MC dynamics for bulk spins. 
Moreover, if the values of the leftmost ($x$ direction) and the 
rightmost spins are fixed to $+1$ 
and $-1$, respectively, an interface perpendicular to the heat flux is generated
between domains with opposite magnetizations. 
If the values of the top ($z$ direction) and bottom spins are fixed to $+1$ and $-1$, 
respectively, 
an interface parallel to the heat flux is formed. In this paper, we investigate both cases.

\section{Thermal conductivity}
In this section we present simulation results for the thermal conductivity,
which is estimated as
\begin{equation}
\kappa(T)=J\frac{L_x}{\Delta T},  \label{eq5}
\end{equation}
where $L_x$ is the system size in the $x$ direction, $J$ heat flux, and $\Delta T=0.05$.
We checked that the result does not seriously change for other choices of $\Delta T$.

First, we deal with the case where the system has an interface perpendicular to
the heat flux.

When we employ the KSC dynamics, a finite-size effect is observed in 
the temperature profile. As seen in Fig.~1, if $L_x$ is small, 
the temperature slope is not uniform and it is larger in higher temperature region.
If $L_x$ is greater than or equal to $64$, the finite-size effect vanishes and the 
uniform temperature gradient is formed.
In the following, large enough $L_x$ is used not to have the bothering finite-size effect.

\begin{figure}[htbp]
\begin{center}
\includegraphics[width=8.6cm]{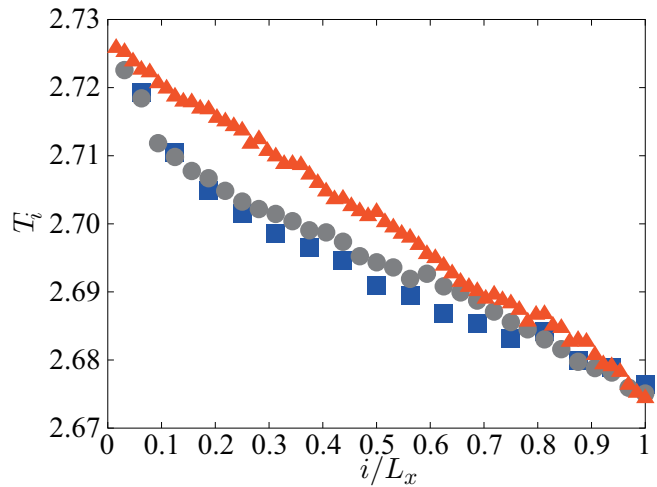}
\caption{Temperature profiles for the system with an interface where the KSC dynamics
is employed. The system size is $L_x \times 16 \times 16, L_x= 16,32,64$ and 
the reservoir temperatures are $T_1=2.725, T_2=2.675$. 
Blue squares, gray circles and orange triangles indicate $L_x=16, 32$ and $64$, respectively.}
%Red, green and blue dots indicate $L_x=16, 32$ and $64$, respectively.}
\label{f2}
\end{center}
\end{figure}

\begin{figure}[htbp]
\centering\includegraphics[width=8.6cm]{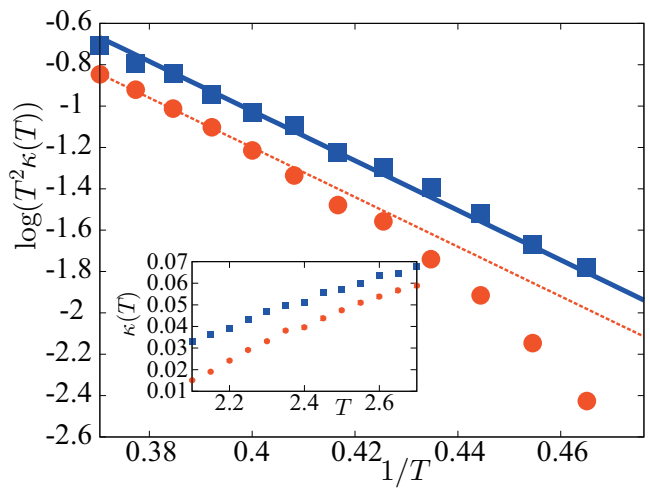}
\caption{For a system of size $64\times 16\times 16$ developed by the KSC dynamics, 
$\log(T^2\kappa(T))$ is shown as a function of $1/T$. 
Inset: plots of $\kappa(T)$ versus $T$. 
Orange circles and blue squares indicate 
the system with and without an interface, respectively.
Orange dotted line and blue line are eye guides to show that $\kappa(T)\sim 1/T^2\exp(-12/T)$ at
the high temperature region.}
\label{f4}
\end{figure}

In Fig.~\ref{f4} we compare the thermal conductivity in the system with and without an
interface. The thermal conductivity is larger in the system without an interface 
than in that with an interface as is the case in the two-dimensional system\cite{6}. 
In $T>T_{\mathrm{R}}^{\mathrm{eq}}$, the thermal conductivity varies like $\kappa(T)\sim\frac{1}{T^2}\exp(-12/T)$ 
in both the cases. This temperature dependence 
is derived from the mean-field-type analysis described in \cite{10}.
In the presence of an interface, the thermal conductivity deviates from the line
below $T_{\mathrm{R}}^{\mathrm{eq}}$, where the variation is more rapid than $\exp(-12/T)/T^2$.
Such a change in the temperature dependence of $\kappa(T)$ is not observed in
the two-dimensional systems\cite{6}.
Thus, we consider the change in temperature dependence of $\kappa(T)$ near $T_{\mathrm{R}}^{\mathrm{eq}}$ 
is an effect from the roughening transition. 
Consider a pair of next-nearest-neighbor spins that are located on the opposite sides 
of a flat interface. To exchange their signs, a large amount of energy is necessary. 
Thus, the exchange of such spins is virtually inhibited in the KS dynamics. 
Hence, energy transport through a smooth surface is very difficult at $T<T_{\mathrm{R}}^{\mathrm{eq}}$. 
This is the reason why the thermal conductivity rapidly decreases below $T_{\mathrm{R}}^{\mathrm{eq}}$.
%As mentioned in \cite{6}, heat transport is very difficult near smooth interface in $T<T_{\mathrm{R}}^{\mathrm{eq}}$ in KS dynamics constituting KSC dynamics. 

\begin{figure}[htbp]
\centering\includegraphics[width=8.6cm]{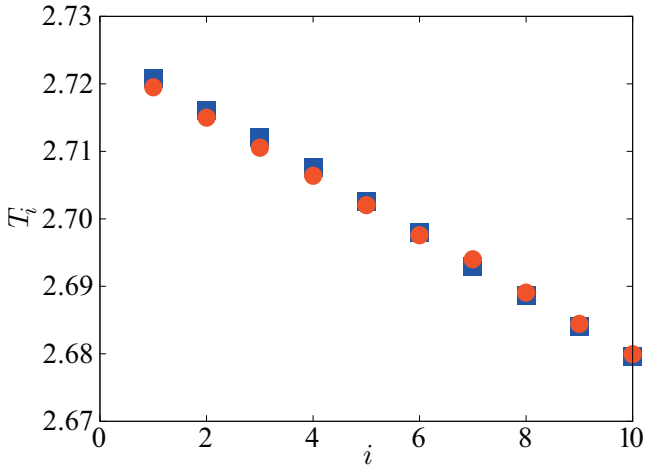}
\caption{Temperature profiles generated by MC dynamics. 
The system size is $10 \times 20 \times 20$ and the reservoir temperatures 
are $T_1 = 2.725$ and $T_2 = 2.675$. Orange circles and blue squares
indicate the system with and without an interface, respectively.}
\label{f5}
\end{figure}

In contrast to the KSC dynamics, the MC dynamics does not suffer from the 
finite-size effect seen in the KSC dynamics as seen in Fig.~\ref{f5}. 
The uniform temperature gradient is realized even in relatively small systems.

\begin{figure}[htbp]
\centering\includegraphics[width=8.6cm]{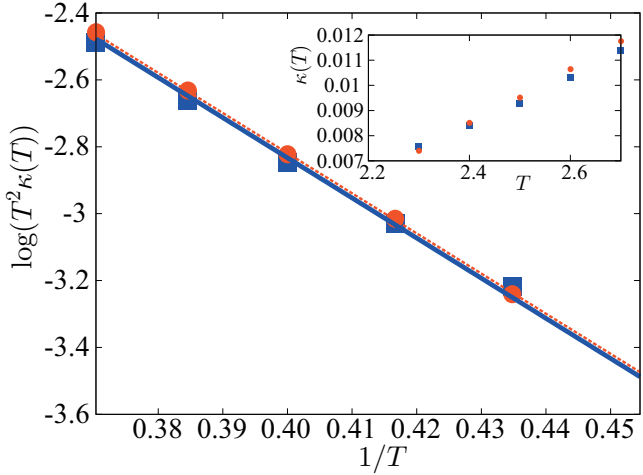}
\caption{Thermal conductivity  $\kappa(T)$ of the system developed by the MC dynamics. 
For a system of size $10\times 20\times 20$, $\log(T^2\kappa(T))$ is plotted as a 
function of $1/T$. Inset: plots of $\kappa(T)$ versus $T$. 
Orange circles and blue squares indicate 
the system with and without an interface, respectively.
Orange dotted line and blue line are eye guides to show that $\kappa(T)\sim 1/T^2\exp(-12/T)$ at
the high temperature region.}
%Green and red dots indicate the system with and without an interface, respectively. 
%Green and red lines are eye guides to show that $\kappa(T)\sim 1/T^2\exp(-12/T)$. }
\label{f7}
\end{figure}

Figure~\ref{f7} show the numerical results of the thermal conductivity in the MC dynamics. 
Contrary to the KSC dynamics, the thermal conductivity is a little bit smaller 
in the system without an interface than in that with an interface. 
Moreover, the thermal conductivity varies like $\kappa(T)\sim\frac{1}{T^2}\exp(-12/T)$
in the whole temperature region. In the MC dynamics, a spin can change its sign 
only with a variation of a single bond energy.  Thus, we consider that 
the flatness of the interface does not affect the thermal conductivity.

\begin{figure}[htbp]
\centering\includegraphics[width=8.6cm]{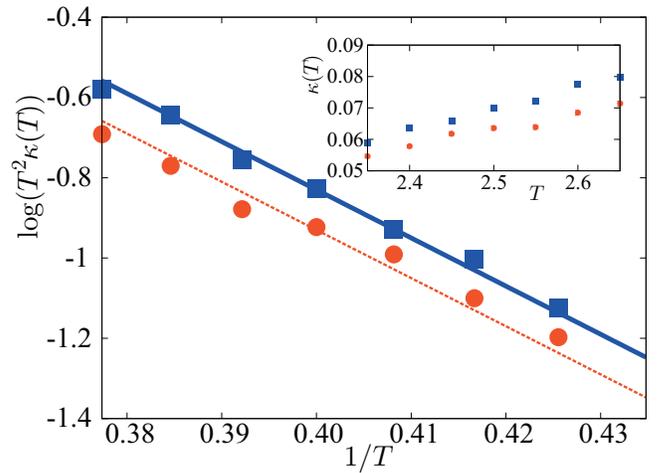}
\caption{Thermal conductivity $\kappa(T)$ of the system developed by the KSC dynamics. 
For a system of size $32\times 32\times 16$, $\log(T^2\kappa(T))$ is shown as a
function of $1/T$. Inset: plots of $\kappa(T)$ versus $T$. 
Orange circles and blue squares indicate 
the system with and without an interface, respectively.
Orange dotted line and blue line are eye guides to show that $\kappa(T)\sim 1/T^2\exp(-12/T)$ at
the high temperature region.}
%Green and red and green dots indicate the system with and without an interface, respectively.
%Green and red lines are eye guides to show that $\kappa(T)\sim 1/T^2\exp(-12/T)$.}
\label{f9}
\end{figure}

\begin{figure}[htbp]
\centering\includegraphics[width=8.6cm]{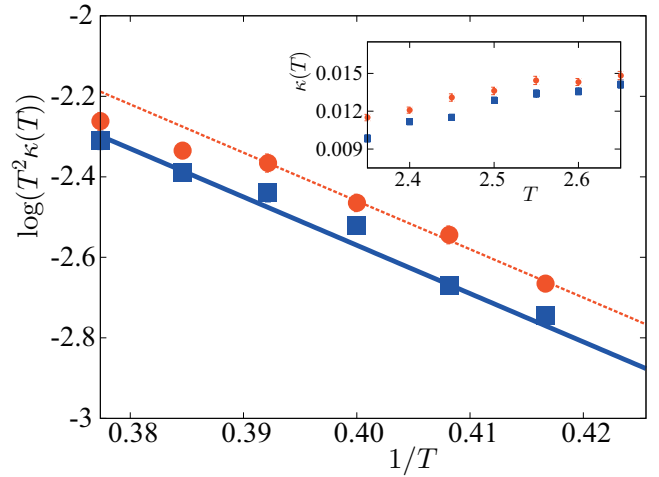}
\caption{Thermal conductivity $\kappa(T)$ of the system developed by the MC dynamics. 
For a system of size $32\times 32\times 16$, $\log(T^2\kappa(T))$ is shown as a function of
 $1/T$. Inset: plots of $\kappa(T)$ versus $T$. 
Orange circles and blue squares indicate 
the system with and without an interface, respectively.
Orange dotted line and blue line are eye guides to show that $\kappa(T)\sim 1/T^2\exp(-12/T)$ at
the high temperature region.}
%Green and red dots indicate the system with and without an interface, respectively. 
%Green and red lines are eye guides to show that $\kappa(T)\sim 1/T^2\exp(-12/T)$.}
\label{f11}
\end{figure}

Now we consider the thermal conductivity when the system has an interface 
parallel to the heat flux using the KSC and MC dynamics. In this case, there is no
noticeable finite-size effects in both the dynamics. 
Moreover, as seen in Figs.~\ref{f9} and \ref{f11} the mean-field type 
temperature dependence can be applied to both the dynamics. 
This is because energy can transport in the region without an interface.

\section{Diffusion constant}

For the interface perpendicular to heat flux, we observe diffusive motion with a drift
to the high-temperature side similar to the two-dimensional case. 
We find that behavior of the diffusion constant of the interface parallel to heat flux
is more interesting. The diffusion constant is estimated as follows.
First, the position of an interface $z$ is defined by using magnetization 
$m=(L_xL_yL_z)^{-1}\sum_{i,j,k}\sigma_{ijk}$ as 
\begin{equation}
z=\frac{L_z(m+m_0)}{2m_0}, \label{eq6}
\end{equation}
where we specified lattice points by the coordinates $(i,j,k)$, and $L_x$, 
$L_y$, and $L_z$ are the system size in each direction, and $m_0$ is spontaneous magnetization.
Thus, if $m=-m_0$, the interface is at the bottom side $z=0$, and if $m=+m_0$,
it is at the top side $z=L_z$. 
The diffusion constant $D$ is calculated from mean square displacements of 
the interface position $z$ as
\begin{equation}
 \langle (z(t)-z(0))^2\rangle=2Dt. \label{eq7}
\end{equation}
Note that temperature varies along an interface in the present setup. 
Thus, the roughness depend on the position on the interface. 
Not with standing that, we can obtain a diffusion 
constant that represents the interface motion as a whole.

\begin{figure}[htbp]
\centering\includegraphics[width=8.6cm]{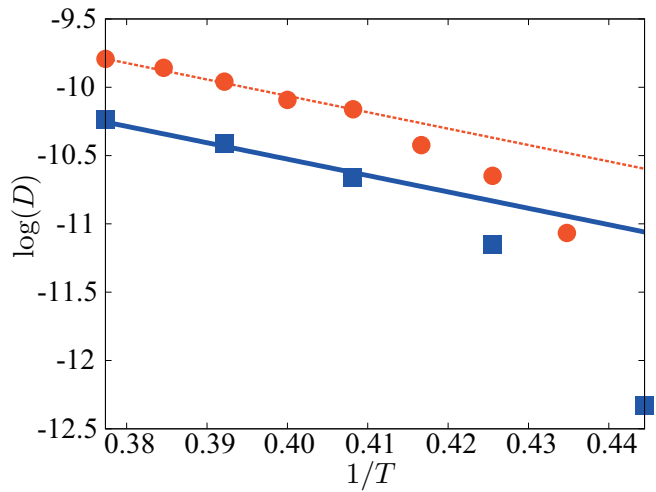}
\caption{For a system of size $32\times 32\times 16$, $\log(D)$ is shown as a function of
$1/T$. Blue squares and orange circles indicate the KSC dynamics and the MC dynamics, respectively. 
Blue and black lines show $D(T)\sim\exp(-12/T)$.}
\label{f12}
\end{figure}

Figure~\ref{f12} shows logarithm of diffusion constant $D$ as a function of $1/T$,
which is obtained by using the KSC and the MC dynamics 
for equilibrium condition $T_1=T_2$. 
The magnitude of the diffusion constant is greater in the MC 
dynamics than in the KSC dynamics. However, the temperature dependence of the
diffusion constant is similar in both the cases. That is, the diffusion constant
is proportional to $\exp(-12/T)$ above $T_{\mathrm{R}}^{\mathrm{eq}}$ and rapidly decreases below $T_{\mathrm{R}}^{\mathrm{eq}}$. 
This result implies that a smooth interface is difficult to move.

\begin{figure}[htbp]
\centering\includegraphics[width=8.6cm]{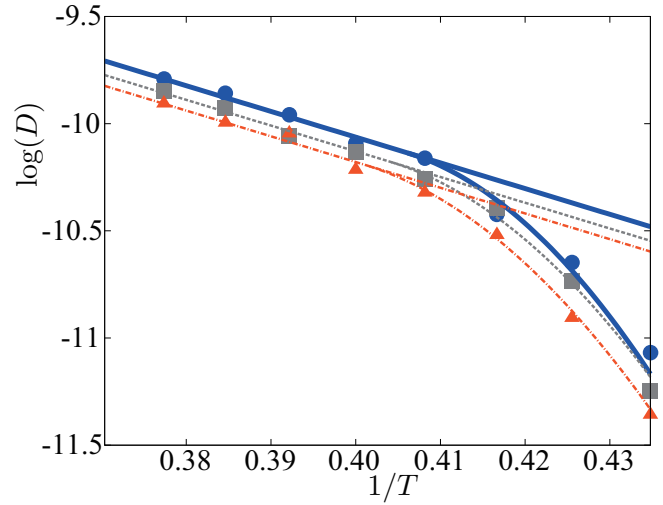}
\caption{For a system of size $32\times 32\times 16$, $\log(D)$ is shown as a
function of $1/T$. Blue circles, gray squares and orange triangles indicate the results of $\Delta T=0.0,0.3$, and
$0.5$, respectively. Blue straight line, gray dotted straight line and orange chain straight line straight lines indicate 
$D_{+}(T)=\exp(-12/T+a)$ and blue curve, gray dotted curve and orange chain curve
indicate $D_{-}(T)=\exp(-12/T+a+A^2(1/T-B)^2)$.}
\label{f13}
\end{figure}

\begin{figure}[htbp]
\centering\includegraphics[width=8.6cm]{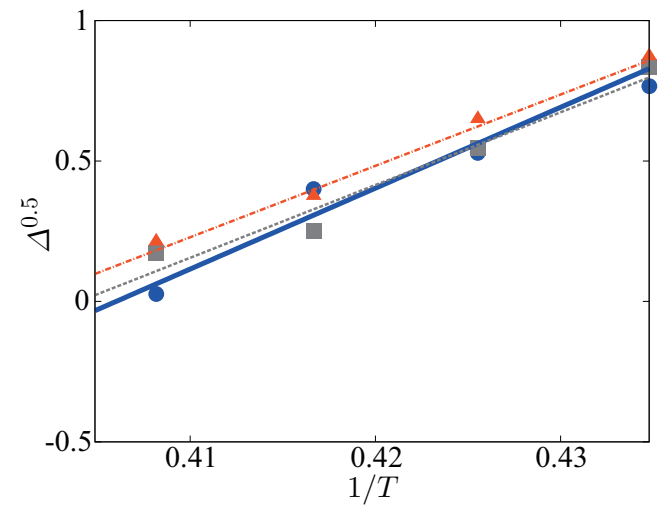}
\caption{Plot of $\sqrt{\Delta}$ versus $1/T$. Blue circles, gray squares and orange triangles indicate 
$\Delta T=0.0,0.3$, and $0.5$, respectively. 
Blue line, gray dotted line and orange chain line are the estimated curves $D_{-}(T)$.}
\label{f15}
\end{figure}

Figure~\ref{f13} shows the logarithm of the diffusion constant $D$ 
obtained from the MC dynamics for the system under temperature gradient. 
At the high-temperature region, the diffusion constant varies with temperature as
$D \sim \exp(-12/T)$ as in the equilibrium case, and
it rapidly decreases below a certain crossover temperature $T_{\mathrm{X}}$. 
We estimate values of $T_{\mathrm{X}}$ in the following manner. 
First, we fit the numerical data by $D_{+}=\exp(-12/T+a)$ in the high-temperature region, 
where $a$ is a fitting parameter. Next, we calculate the deviation from $D_{+}$ 
as $\Delta=\log D_{+}-\log D$. 
As seen in Fig.~\ref{f15}, $\sqrt{\Delta}$ is roughly proportional to $1/T$ in 
the low-temperature region, where we fit the data by $\sqrt{\Delta}=A(1/T-B)$
with parameters $A$ and $B$. 
That is, the diffusion constant in low-temperature region is fitted by 
$D_{-}=\exp(-12/T+a-A^2(1/T-B)^2)$ as seen in Fig.~\ref{f13}.
Then we identify the crossover temperature as $T_{\mathrm{X}}=1/B$.
The obtained crossover temperature shows temperature dependence like 
$T_{\mathrm{X}} \sim T_{\mathrm{R}}^{\mathrm{eq}}+0.127\Delta T$ as in Fig.~\ref{f14}.

In equilibrium system, $T_{\mathrm{X}}$ and $T_{\mathrm{R}}^{\mathrm{eq}}$ are indistinguishable.
Thus, the above result implies the possibility that the roughening transition temperature is 
shifted by heat conduction. To verify the implication, we estimate the roughening
transition temperature in the system with heat conduction by using 
the width $W$ of an interface defined as\cite{18}
\begin{equation}
W^2= \frac{1}{(L_xL_y)^2}\sum_{i,j,k,l}\langle (h_{ij}-h_{kl})^2 \rangle, \label{eq16}
\end{equation}
where $h_{ij}=1/(2m_0)\sum_{k} \sigma_{ijk}$ is the height of the interface at $(x,y)=(i,j)$.
It is known that in equilibrium $W^2$ behaves as follows\cite{8}
\begin{align}
W^2&\sim (c_1+c_2(T-T_{\mathrm{R}}^{\mathrm{eq}})^{1/2})\log L &(T > T_{\mathrm{R}}^{\mathrm{eq}})\label{eq17}\\
W^2&\sim c_3+c_4(T_{\mathrm{R}}^{\mathrm{eq}}-T)^{-1/2} & (T < T_{\mathrm{R}}^{\mathrm{eq}}), \label{eq18}
\end{align}
where $c_1$, $c_2$, $c_3$ and $c_4$ are some constants.

\begin{figure}[htbp]
\centering\includegraphics[width=8.6cm]{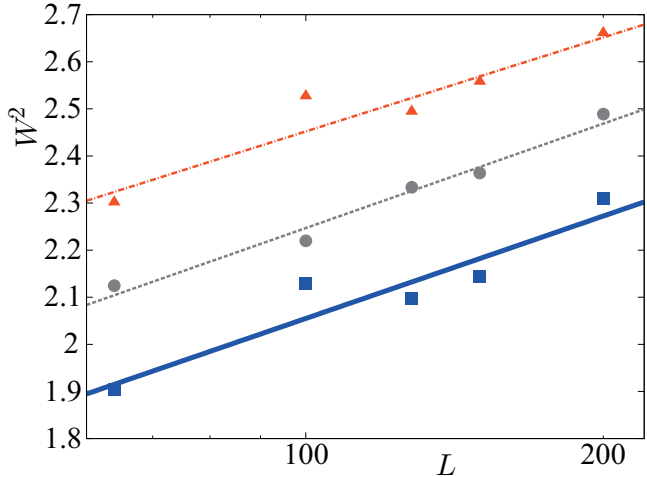}
\caption{Size dependence of $W^2$ for the high temperature region $T>T_{\mathrm{R}}^{\mathrm{eq}}$ for
the system of size $L\times L\times 20$ and the temperature difference $\Delta T=0.1$.
Blue circles, gray squares and orange triangles indicate $T=2.55,2.60$, and $2.65$, respectively. 
Blue line , gray dotted line and orange chain line show $W^2\sim \log L$.}
\label{f17}
\end{figure}

\begin{figure}[htbp]
\centering\includegraphics[width=8.6cm]{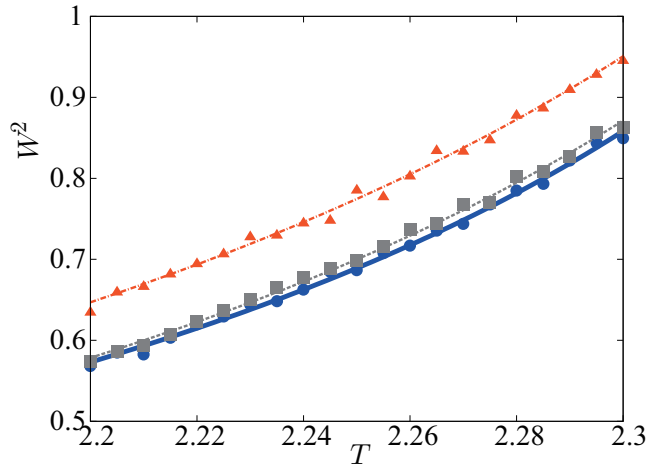}
\caption{Temperature dependence of $W^2$ for the low temperature region $T<T_{\mathrm{R}}^{\mathrm{eq}}$
for the system of size $128\times128\times20$.
Blue circles, gray squares and orange triangles indicate $\Delta T=0.0,0.1$, and $0.3$, respectively. 
Blue, black and orange lines show Eq. \ref{eq18}.
}
\label{f16}
\end{figure}

\begin{figure}[htbp]
\centering\includegraphics[width=8.6cm]{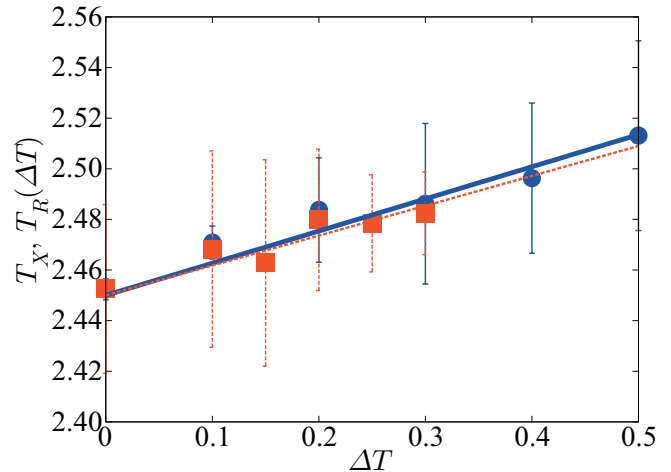}
\caption{$T_{\mathrm{X}}$ (blue circles) and $T_{\mathrm{R}}(\Delta T)$ (orange squares).
%Blue dots indicate the results in the case where spontaneous magnetization is not considered. Black dots indicate the results in the case where spontaneous magnetization is considered. 
%Inset: Comparison between $T_{\mathrm{R}}(\Delta T)$ by the MC dynamics and that by the Glauber dynamics (orange triangles). 
Blue, black and orange lines are fitted straight lines.}
\label{f14}
\end{figure}

We numerically calculated $W^2$ for the systems with a fixed temperature difference 
$\Delta T$, various average temperature $T$ and various system size $L$.
As the result we found that there is a temperature $T_{\mathrm{R}}(\Delta T)$ such
that if $T>T_{\mathrm{R}}(\Delta T)$, $W^2$ behaves like Eq.~(\ref{eq17}) (Fig.~\ref{f17})and 
if $T>T_{\mathrm{R}}(\Delta T)$, Eq.~(\ref{eq18}) is well satisfied (Fig.~\ref{f16}).
Thus we call $T_{\mathrm{R}}(\Delta T)$ the nonequilibirum 
roughening transition temperature. Note that we always write the argument $\Delta T$
to distinguish it from the equilibrium roughening transition temperature 
$T_{\mathrm{R}}^{\mathrm{eq}}$.

The nonequilibrium roughening transition temperature 
$T_{\mathrm{R}}(\Delta T)$ thus obtained varies with 
$\Delta T$ like $T_{\mathrm{R}}(\Delta T) \sim T_{\mathrm{R}}^{\mathrm{eq}}+0.118\Delta T$.
Figure~\ref{f14} shows comparison between $T_{\mathrm{X}}$ and $T_{\mathrm{R}}(\Delta T)$,
which shows that $T_{\mathrm{X}}$ and $T_{\mathrm{R}}(\Delta T)$ agree with each other within error bars.

%\textcolor{red}{Furthermore, we can define the position of the interface $z'$ and the height of the interface $h'_{i,j}$ taking spontaneous magnetization $m_0$ into account as }

%\begin{equation}
%\textcolor{red}{z'=\frac{L_z}{2}(m/m_0+1),} \label{eq6_1}
%\end{equation}
%\begin{equation}
%\textcolor{red}{h'_{i,j}=\frac{1}{2m_0}\sum_{k}\sigma_{ijk}.} \label{eq7_1}
%\end{equation}
%\textcolor{red}{The temperature dependences of the diffusion constant $D$ and the width of the interface $W$ by these definition are similar to them by the previous definition.
%The crossover temperature $T'_X$ and the roughening transition temperature $T'_R(\Delta T)$ estimated from these difinition are shown in Fig.~\ref{f14}. $T'_X$ and $T'_R(\Delta T)$ vary with $\Delta T$ like $T'_X=T_{\mathrm{R}}^{\mathrm{eq}}+0.127\Delta T $ and $T'_R(\Delta T)=T_{\mathrm{R}}^{\mathrm{eq}}+0.125\Delta T$. These are more consistant than $T_{\mathrm{R}}(\Delta T)$.  Thus, our numerical results can be regarded as an evidence of the shift of the roughening transition temperature by heat conduction. }

%\textcolor{red}{We also calculate the width of interface $W$ by the Glauber dynamics on the basis of the temperature profile by the  MC dynamics. The roughening temperature $T_{\mathrm{R}}(\Delta T)$ estimated from the result is shown in the inset of Fig.\ref{f14} and  varies with $\Delta T$ like $T'_R(\Delta T)=T_{\mathrm{R}}^{\mathrm{eq}}+0.152\Delta T $ .  The shift of the roughening transition temperature is also confirmed in the Glauber dynamics.}

\section{Summary and Discussion}
In this paper, we have numerically studied the relationship between the diffusion of an interface and heat conduction in the three-dimensional Ising model. We have examined how the dynamics and the arrangements of an interface affect heat conduction and the interface motion.

First, we investigated heat conduction in the two cases 
where the interface is perpendicular and parallel to the heat flux 
and with two kinds of dynamics; the KSC dynamics and the MC dynamics. 
We have found that whether an interface enhances heat conduction or not 
depends on dynamics. It is the case in the MC dynamics, but it is not in the KSC dynamics. 
In the case of an interface perpendicular to heat flux, 
the KSC dynamics yields a sudden decrease of thermal conductivity 
just below $T_{\mathrm{R}}^{\mathrm{eq}}$, 
while the MC dynamics does not. 
It shows that the MC dynamics is superior to the KSC dynamics for the use 
in low-temperature simulations.  

Next, we computed the diffusion constant in the case where the interface is parallel to 
heat flux. The diffusion constant showed crossover in temperature dependence 
irrespective of dynamics. We estimated the crossover temperature $T_{\mathrm{X}}$, 
which agrees with the roughening transition temperature $T_{\mathrm{R}}^{\mathrm{eq}}$ in equilibrium and 
deviates from it in the presence of temperature gradient. It suggests some relationship 
between the roughness and the motion of the interface, but the functional form 
used for fitting is ad hoc and lacks any theoretical grounds. 
Then, we calculated the width of the interface in the systems 
with a boundary-temperature difference $\Delta T$ and determined 
the nonequilibrium roughening transition temperature from 
their dependence on system size and temperature. 
The obtained nonequilibrium roughening transition temperature 
$T_{\mathrm{R}}(\Delta T)$ agrees with $T_{\mathrm{X}}$ within error bars, 
though the data is rather noisy. One may suspect that
the result depends on dynamics. We carried out simulations with the Glauber 
dynamics with the same temperature profile as obtained in the MC dynamics and
obtained almost the same result. Thus we do not consider that the behavior of 
$T_{\mathrm{X}}$ and $T_{\mathrm{R}}(\Delta T)$ come
from the peculiarity of the dynamics employed.
The above results suggest the conjecture 
that heat conduction shifts the roughening-transition temperature. 
To our knowledge, it is the first time that such evidence is found for the motion of 
the interface motion in the Ising model. 

%\textcolor{red}{
%In this study, because we consider heat conduction situation, deterministic dynamics that relaxation is slow are used. Thus the computation time of the system is very long, so we cannot performed the computation in a large system. In future, large size numerical calculation using GPUs should be performed.}
%The system sizes considered in this study is moderate. In the future, numerical calculation with large system size using GPUs should be executed.

To establish this conjecture, we have to improve computational performance and 
develop theoretical considerations.  The dynamics we employed in this study 
conserves local energy. In contast to usual Monte Carlo dynamics, it is not as easy 
to accelerate or parallelize such dynamics. 
Thus we have been limited to modest system sizes. 
Improvements using, for example, the GPU are a future problem.
In the classification by Hohenberg and Halperin\cite{19}, 
energy-conserving and magnetization-nonconserving dynamics like the KSC and MC 
dynamics are classified as Model C. 
A theoretical study of our findings based on Model C is desirable,
because it means that the phenomena have a universal feature.

\section*{Acknowledgements}
Numerical computation in this work was carried out at the 
Yukawa Institute Computer Facility.
%\newpage

\end{document}